\documentclass[aps, pra, 10pt, amsmath, twocolumn, superscriptaddress, floatfix,longbibliography]{revtex4-1}
\usepackage[english]{babel}
\usepackage{graphicx}
\usepackage[charter]{mathdesign}
\usepackage{bm}
\usepackage[svgnames]{xcolor}
\newcommand\kv{\mathbf{k}}
\newcommand\gv{\mathbf{g}}
\newcommand\dv{\mathbf{d}}
\newcommand\ev{\mathbf{e}}
\newcommand\xv{\mathbf{x}}
\newcommand\yv{\mathbf{y}}
\newcommand\rv{\mathbf{r}}
\newcommand\Kv{\mathbf{K}}
\newcommand\Gv{\mathbf{G}}
\newcommand\kvt{\mathbf{\tilde k}}
\newcommand\sigmav{\bm{\sigma}}
\newcommand\Sigmav{\bm{\Sigma}}
\newcommand{\fdag}{{\phantom{\dagger}}}
\renewcommand{\c}{c^\fdag}
\newcommand{\cd}{c^\dagger}
\newcommand{\up}{{\uparrow}}
\newcommand{\down}{\downarrow}
\newcommand\ket[1]{|#1\rangle}
\newcommand\bra[1]{\langle#1|}

\begin{document}
\title{Parity-mixing superconducting phase in the Rashba-Hubbard model and its topological properties from dynamical mean field theory}
\author{Xiancong Lu}\affiliation{Department of Physics, Xiamen University, Xiamen 361005, China} 
\author{D. S\'en\'echal}\affiliation{D\'epartement de physique and Institut quantique, Universit\'e de Sherbrooke, Sherbrooke, Qu\'ebec, Canada J1K 2R1} 
\begin{abstract}
We investigate parity-mixing superconductivity in the two-dimensional Hubbard model with Rashba spin-orbit coupling, using Cellular Dynamical Mean-Field Theory (CDMFT).
A superconducting state with mixed singlet $d$-wave and triplet $p$-wave character is found in a wide range of doping. 
The singlet component decreases with the amplitude of the Rashba spin-orbit coupling, whereas the triplet component increases, but both disappear at about 20\% doping.
The effect of a Zeeman field is also investigated; it tends to suppress both types of superconductivity, but induces nontrivial topological properties: the computed bulk Chern number is nonzero in the mixed superconductivity phase, at least in the underdoped region.
A strong suppression of the excitation gap occurs slightly after optimal doping; this might be the sign of a topological transition within the superconducting dome.
\end{abstract}
\maketitle
\section{Introduction}
Superconductors with nontrivial topology are the focus of attention because of their ability to host Majorana zero modes, which obey non-Abelian statistics and are relevant to topological quantum computation~\cite{qi.zh.11, sa.an.17}.
Many theoretical avenues for the realization of topological superconductivity (TSC) have been put forward~\cite{sa.an.17}. 
An early promising proposal towards TSC was to use an odd-parity superconducting state~\cite{re.gr.00, kita.01, ivan.01}.
Later, a conventional even-parity pairing state was also shown to be a possible route towards TSC, for example, on the surface of a topological insulator~\cite{fu.ka.08}, or in superconductors or semiconductors with Rashba spin-orbit coupling~\cite{sa.ta.09, sa.fu.10,sa.lu.10, alic.10}.
In particular, Sato et al. demonstrated that a nodal $d$-wave superconductor with Rashba spin-orbit coupling is a ``weak'' TSC in the presence of a Zeeman magnetic field~\cite{sa.fu.10}. 
Along this direction, it was observed recently that a mixed $d$- and $p$-wave Rashba superconductor is \textit{gapful} in a magnetic field, and therefore is a ``strong'' TSC with nontrivial Chern number~\cite{yo.ya.16, da.ya.16}. 
One advantage of this scenario towards TSC is that no fine-tuning of parameters is needed to realize the topological order. 
It is believed that parity-mixing Cooper pairs between singlet and triplet states occur in noncentrosymmetric superconductors with broken inversion symmetry (e.g., asymmetric spin-orbit coupling) such as $\rm CePt_3Si$~\cite{fr.ag.04, ba.si.12, ta.ka.09}. 
Therefore, according to the above scenario, TSC could be realized intrinsically in natural solid-state systems. 
However, previous studies~\cite{sa.ta.09,sa.fu.10,yo.ya.16, da.ya.16} are mainly based on Bogoliubov-de Gennes
(BdG) mean-field Hamiltonians, in which superconductivity is assumed to exist. 
The effect of interaction between electrons is not fully addressed; neither is the effect of spin-orbit coupling and Zeeman magnetic field on the superconducting order parameters.
In this work, we study the Hubbard model with Rashba spin-orbit coupling and Zeeman magnetic field, in order to probe the stability of superconducting state in a variety of parameter regions.
We use cluster dynamical mean-field theory with an exact diagonalization impurity solver in order to see the superconducting state emerge naturally from strong electron-electron repulsion. 
The paper is organized as follows:
In Sec.~\ref{sec:intro}, we introduce the model Hamiltonian and the numerical method used.
In Sec.~\ref{sec:topo}, we explain how to extract topological properties, in particular the Chern invariant.
In Sec.~\ref{sec:res}, we present the main results: the parity-mixing superconducting phase as a function of hole doping for several values of the Rashba coupling and of the Zeeman field. The solutions we find have nontrivial topology when the Zeeman field is nonzero, but this topological character essentially disappears when doping exceeds optimal doping, before the order parameter actually vanishes.
We end with a conclusion and perspectives.
\section{Model and method}\label{sec:intro}
\subsection{Noninteracting Hamiltonian}
The one-band Rashba-Hubbard model on a two-dimensional (2D) square lattice is defined by the following Hamiltonian, in the absence of interactions:
\begin{align}\label{eq:H0k}
H_0 &= \sum_\kv  \cd_\kv \mathcal{H}^0_\kv \c_\kv \notag \\
&= \sum_\kv  \cd_\kv\left(\epsilon_\kv - h\sigma_3 + 2\lambda \sigmav\cdot\gv_\kv \right)\c_\kv
\end{align}
where $\c_\kv = (\c_{\kv\up}, \c_{\kv\down})$ is a doublet of annihilation operators for the two spin projections and wave vector $\kv$, $h$ is an external Zeeman field, $\sigmav$ is the vector of Pauli matrices, and 
\begin{align}
  \epsilon_\kv &= -2t(\cos k_x + \cos k_y) - \mu \notag \\
  g_\kv &= (-\sin k_y, \sin k_x, 0)
\end{align}
The $2\times2$ matrix $\mathcal{H}^0_\kv$ is the momentum-dependent, noninteracting Hamiltonian.
The hopping amplitude between nearest-neighbor sites is set to $t=1$ in the remainder of this paper;
we include the chemical potential $\mu$ into the dispersion relation.
$\lambda$ is the Rashba spin-orbital coupling.
Note that $\gv_\kv = -\gv_{-\kv}$, and thus the Rashba coupling breaks inversion symmetry, but does not break time-reversal invariance; the Zeeman field $h$ does, however.
This is the model adopted by Yoshida and Yanase~\cite{yo.ya.16}.
In the presence of superconductivity, the mean-field, Bogoliubov-de Gennes Hamiltonian requires a doubling of the degrees of freedom in order to be expressed in the usual form:
\begin{equation}
H^{\rm BdG} = \frac12 \sum_\kv \Psi_\kv^\dagger \mathcal{H}_\kv^{\rm BdG} \Psi_\kv
\end{equation}
where we introduced the four-component annihilation operator $\Psi_\kv = \left(\c_{\kv\up}, \c_{\kv\down}, \cd_{-\kv\up}, \cd_{-\kv\down}\right)$.
The momentum-dependent $4\times4$ matrix  $\mathcal{H}_\kv^{\rm BdG}$ takes the form
\begin{equation}
  \mathcal{H}_\kv^{\rm BdG} = 
  \begin{pmatrix}
    \mathcal{H}^0_\kv & \Delta_\kv \\ \Delta_\kv^\dagger & -\mathcal{H}_{-\kv}^{0T}
  \end{pmatrix}
\end{equation}
where the  $2\times2$ matrix $\Delta_\kv$ is the gap function:
\begin{equation}
  \Delta_\kv = i\left( \psi_\kv + \dv_\kv\cdot\sigmav\right)\sigma_2
\end{equation}
The singlet gap function $\psi_\kv$ is here combined with the triplet gap function, represented by the vector $\dv_\kv$.
Singlet, $d_{x^2-y^2}$-wave superconductivity is characterized by the gap function $\psi_\kv = \Delta_d(\cos k_x-\cos k_y)$.
On the other hand, a greater variety of triplet states is possible.
We will probe the state studied by Yoshida and Yanase~\cite{yo.ya.16}, in which the $d$-vector takes the form
$\dv_\kv = \Delta_p(\sin k_y, \sin k_x, 0)$.
This triplet gap function is chosen because it belongs to the same irreducible representation of the point group $C_{4v}$ of the noninteracting Hamiltonian as the $d$-wave gap function, and thus can mix with it.
Note that the $d$-vector is not parallel to the $\gv$ vector, contrary to the BCS-like solution investigated by Frigeri et al.~\cite{fr.ag.04}.
One could also define a chiral $p_x+ip_y$ state where $\dv_\kv = \Delta_p(0, 0, \sin k_x + i\sin k_y)$, but this possibility is not investigated here.
\subsection{Real-space representation Hamiltonian}
The focus of our work is the Rashba-Hubbard model, obtained from Hamiltonian~\eqref{eq:H0k} by adding the on-site, screened Coulomb interaction $U$:
\begin{equation}\label{RH}
  H^{RH} = H_0 + U\sum_\rv n_{\rv\uparrow}n_{\rv\downarrow}
\end{equation}
where $n_{\rv\sigma}$ is the number of electrons of spin $\sigma$ at site $\rv$.
In the real-space representation, more practical within our methodology, the noninteracting Hamiltonian takes the following form:
\begin{align}
&H_0 = - t\sum_{\rv\sigma} \sum_{\ev=\xv,\yv}
        ( \cd_{\rv\sigma} \c_{\rv+\ev,\sigma} + \mbox{H.c.}) 
      \notag \\
   & -\lambda \sum_\rv \Big[ 
        (\cd_{\rv-\xv\down} \c_{\rv\up} - \cd_{\rv+\xv\down} \c_{\rv\up})
   + i(\cd_{\rv-\yv\down} \c_{\rv\up} - \cd_{\rv+\yv\down} \c_{\rv\up})
      + \mbox{H.c.} \Big]\notag \\
   & - h\sum_\rv (\cd_{\rv\up} \c_{\rv\up} - \cd_{\rv\down} \c_{\rv\down})
      - \mu\sum_{\rv\sigma} n_{\rv\sigma}.
\end{align}
Here $\xv$ and $\yv$ are the lattice unit vector along the $x$ and $y$ direction.
Model~(\ref{RH}) can be realized in electronic systems such as the (001) interface of superconductors $\rm Sr_2RuO_4$~\cite{ta.ka.09} and the high-$T_c$ cuprates thin film on a substrate with asymmetric potential~\cite{yo.ya.16} irradiated by circularly polarized laser light~\cite{ta.da.17}.
It also can be implemented in optical lattices of ultracold atoms, in which the 2D spin-orbit coupling and topological bands are realized through an optical Raman lattice~\cite{wu.zh.16}.
\begin{figure}[h]
  \includegraphics[width=0.7\columnwidth]{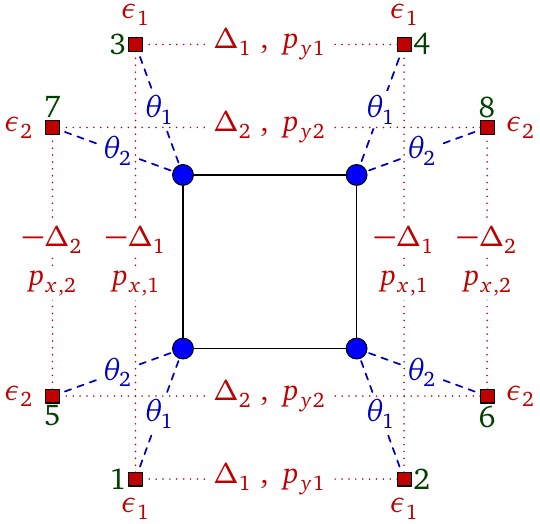}
  \caption{(Color online). The cluster-bath systems used in our implementation of ED-CDMFT. See text for details.\label{fig:cluster_bath}}
\end{figure}
  
\subsection{Impurity model}
In order to study the possible superconducting state in model~\eqref{RH}, we use Cellular Dynamical Mean-Field Theory (CDMFT)~\cite{li.ka.00,ko.sa.01,li.is.08,sene.15} with an exact diagonalization solver at zero temperature (or ED-CDMFT).
In CDMFT, the infinite lattice is tiled into small clusters, each of which is then coupled to a bath of uncorrelated, auxiliary orbitals. 
We will not review the CDMFT procedure here, but rather refer to the appropriate literature.
Let us nevertheless recall that the cluster, along with its bath environment, is described by an \textit{Anderson impurity model} (AIM), defined by the following Hamiltonian, in the absence of superconductivity:
\begin{equation}\label{eq:Himp}
H_{\rm imp} = H^{RH}_c + \sum_{i,r} \theta_{ir} \left(c_{i\sigma}^\dag a_{r\sigma}^\fdag + \mbox{H.c.} \right)
+ \sum_r \epsilon_r a_{r\sigma}^\dag a_{r\sigma}^\fdag,
\end{equation}
where $H^{RH}_c$ is the Hamiltonian~\eqref{RH}, but restricted to the cluster; $a_{r\sigma}$ annihilates an electron with spin $\sigma$ on the bath orbital labelled by $r$, $\theta_{ir}$ is a
hybridization parameter between cluster site $i$ and orbital $r$, and $\epsilon_r$ is a bath energy.
The bath parameters $\theta_{ir}$ and $\epsilon_r$ are determined by an approximate self-consistent procedure, as proposed initially in~\cite{Caffarel1994}.
We use a four-site ($2\times2$) cluster-bath system, as illustrated on Fig.~\ref{fig:cluster_bath}.
Each cluster site is associated with two baths orbitals.
In order to probe superconductivity, we further add singlet and triplet pairing operators within the bath, in addition to Hamiltonian~\eqref{eq:Himp}.
Given two sites labelled by $i$ and $j$, the following pairing operators may be defined:
\begin{align}
  &~~\hat\Delta_{ij} = a_{i\up}a_{j\down}-a_{i\down}a_{j\up} &\mbox{(singlet)} \\
  &\left.\begin{aligned}
    \hat{d}_{ij}^{(x)} &= a_{i\up} a_{j\up} - a_{i\down} a_{j\down} \\ 
    \hat{d}_{ij}^{(y)} &= i\left(a_{i\up} a_{j\up} + a_{i\down} a_{j\down}\right)  \\
    \hat{d}_{ij}^{(z)} &= \left(a_{i\up} a_{j\down} +  a_{i\down} a_{j\up}\right) 
  \end{aligned}\quad \right\}&\mbox{(triplet)}
\end{align}
These pairing terms are added to the bath Hamiltonian, in order to allow the system to be spontaneously pushed towards superconductivity within the DMFT self-consistency procedure.
The cluster Hamiltonian, however, will not contain pairing terms, even though various SC order parameters will be measured from the anomalous Green function derived from the impurity problem (see below).
In terms of the bath orbitals numbering scheme illustrated in Fig.~\ref{fig:cluster_bath}, the pairing terms added to the bath Hamiltonian are
\begin{align}\label{eq:bath_params}
H_{\rm sc} &= \Delta_1\left(\hat\Delta_{12} + \hat\Delta_{34} - \hat\Delta_{13} - \hat\Delta_{24}\right) \notag \\
&+ \Delta_2\left(\hat\Delta_{56} + \hat\Delta_{78} - \hat\Delta_{57} - \hat\Delta_{68}\right) \notag \\
&+ ip_{x1}\left(\hat d^{(y)}_{12} + \hat d^{(y)}_{34}\right) + ip_{y1}\left(\hat d^{(x)}_{13} + \hat d^{(y)}_{24}\right) \notag \\
&+ ip_{x2}\left(\hat d^{(y)}_{56} + \hat d^{(y)}_{78}\right) + ip_{y2}\left(\hat d^{(x)}_{57} + \hat d^{(y)}_{68}\right)
+ \mbox{H.c.}
\end{align}
In the restricted scheme used here, the AIM is characterized by 10 variational parameters, all illustrated on Fig.~\ref{fig:cluster_bath}: bath levels $\epsilon_{1,2}$, hybridization amplitudes $\theta_{1,2}$, in-bath singlet pairing amplitudes $\Delta_{1,2}$ and in-bath triplet pairing amplitudes $p_{x1}$, $p_{x2}$, $p_{y1}$, $p_{y2}$.
Trial values are given to these parameters, and the DMFT procedure leads to converged values, which also yield a Green function (and associated self-energy) for the $2\times2$ plaquette.
Each step in the ED-CDMFT procedure requires the solution, via the Lanczos method and its variants, of the impurity model, within a Hilbert space of dimension $2^{24}\sim 1,6\times 10^7$.
The self-energy is then extracted and applied to the whole lattice.
This, in essence, is the ED-CDMFT method.
Once a solution is found for a given set of model parameters ($U$, $\mu$, $h$, $\lambda$, etc.), average values of one-body operators defined on the lattice can be computed from the Green function.
In particular, we are interested in the singlet and triplet superconducting order parameters, which can be defined as follows:
\begin{align}\label{eq:OP}
  \hat{D} &= \frac1N\sum_{\rv}  \left(\hat\Delta_{\rv, \rv+\xv}-\hat\Delta_{\rv, \rv+\yv}\right) \\
  \hat{p}_x &= \frac1N\sum_{\rv}  \hat d^{(y)}_{\rv, \rv+\xv} \\
  \hat{p}_y &= \frac1N\sum_{\rv}  \hat d^{(x)}_{\rv, \rv+\yv}
\end{align}
($N$ is the number of lattice sites).
Let us stress that the above operators are not imposed on the Hamiltonian; only their expectation value is measured, using the Green function obtained from ED-CDMFT.
That Green function is computed from the cluster's self-energy $\Sigmav(\omega)$ as
\begin{equation}\label{eq:latticeGF}
\Gv^{-1}(\kvt,\omega) = \Gv_0^{-1}(\kvt,\omega) - \Sigmav(\omega)
\end{equation}
Here $\kvt$ denotes a reduced wave vector, belonging to the Brillouin zone associated with the superlattice of plaquettes that defines a tiling of the square lattice, and $\Gv_0$ is the noninteracting Green function.
This is a mixed representation, in which the plaquette is treated in a real-space representation whereas the superlattice of plaquettes is treated in reciprocal space. Thus the matrices involved ($\Gv$, $\Gv_0$,$\Sigmav$) are $4N_c\times 4N_c$, where $N_c=4$ is the number of sites in the plaquette and the other factor of 4 comes from spin and Nambu doubling.
We can go back to a fully wave vector-dependent representation $\mathcal{G}(\kv,\omega)$, where $\kv$ now belongs to the original Brillouin zone and $\mathcal{G}$ is a smaller, $4\times4$ matrix by a procedure called \textit{periodization}.
The simplest periodization scheme is to Fourier transform $\Gv$ directly from the plaquette to the original Brillouin zone, as follows~\cite{Senechal:2000}:
\begin{equation}\label{eq:Gper}
\mathcal{G}_{ij}(\kv,\omega) = \frac1{N_c} \sum_{\rv, \rv'} e^{-i\kv\cdot(\rv-\rv')}G_{\rv i, \rv' j}(\kv, \omega)
\end{equation}
where $i,j$ are spin and Nambu indices, and the difference between $\kv$ and $\kvt$ is an element $\Kv$ of the reciprocal superlattice: $\kv = \kvt+\Kv$. Note that $\Gv(\kv, \omega)=\Gv(\kvt, \omega)$ since $\Gv$ is by construction a periodic function of the reduced Brillouin zone.
\section{Topological properties}\label{sec:topo}
In this section we explain how to compute the Chern number associated with a solution found by CDMFT.
The evaluation of topological invariants for interacting systems is much more difficult than for noninteracting systems. 
Several schemes based on the single-particle Green function have been proposed~\cite{is.ma.86, volo.03, wa.qi.10, wa.zh.12a}.
We will adopt the one proposed by Wang \textit{et al.}~\cite{wa.zh.12a, wa.zh.12b} for the interacting topological superconductors, in which the topological invariants are constructed from the single-particle Green function at zero frequency.
The key idea is to define a fictitious, noninteracting ``topological Hamiltonian'' $h_t(\kv)$ from the inverse of the zero-frequency Green function:
\begin{equation} 
h_t(\kv) = - \mathcal{G}^{-1}(\kv,\omega=0), 
\end{equation}
This Hamiltonian can be diagonalized:
\begin{equation}
h_t(\kv) \ket{\alpha,\kv} = \mu_\alpha (\kv) \ket{\alpha,\kv}
\end{equation}
with eigenvalues $\mu_\alpha(\kv)$ and eigenvectors $\ket{\alpha,\kv}$.
One can then define a generalized Chern number just like in noninteracting systems~\cite{wa.zh.12a,wa.zh.12b}:
\begin{equation}\label{eq:Chern}
C_1 = \int \frac{d^2k}{2\pi} \; \mathcal{F}_{xy} (\kv)
\quad\quad
\mathcal{F}_{xy} (\kv) = \frac{\partial \mathcal{A}_y}{\partial k_x} - \frac{\partial \mathcal{A}_x}{\partial k_y}
\end{equation}
where the Berry connection is 
\begin{equation}
\mathcal{A}_j (\kv) = -i \sum_{\mu_\alpha(\kv)<0} \bra{\alpha, \kv} \partial_{k_j} \ket{\alpha,\kv},
\qquad (j = x,y)
\end{equation}
The $4\times4$ matrix $-\mathcal{G}^{-1}(\kv,\omega=0)$ can be diagonalized to obtain four bands, labeled $\mu_\alpha(\kv)$, ($\alpha=1,\dots,4$).
Because of particle-hole symmetry in superconductors, the following relations hold:
\begin{equation}\label{eq:spectrumG}
\mu_3(\kv)=-\mu_2(-\kv)\qquad\mu_4(\kv)=-\mu_1(-\kv)
\end{equation}
with $\mu_{1,2}(\kv)\leq0$. 
The negative energy states are fully occupied as in an insulating state.
Of course, these are not really one-particle states of the system: they are fictitious noninteracting states of the topological Hamiltonian $h_t(\kv)$ whose only purpose is to allow for the computation of the Chern invariant.
\begin{figure}[h]
\includegraphics[width=0.94\columnwidth]{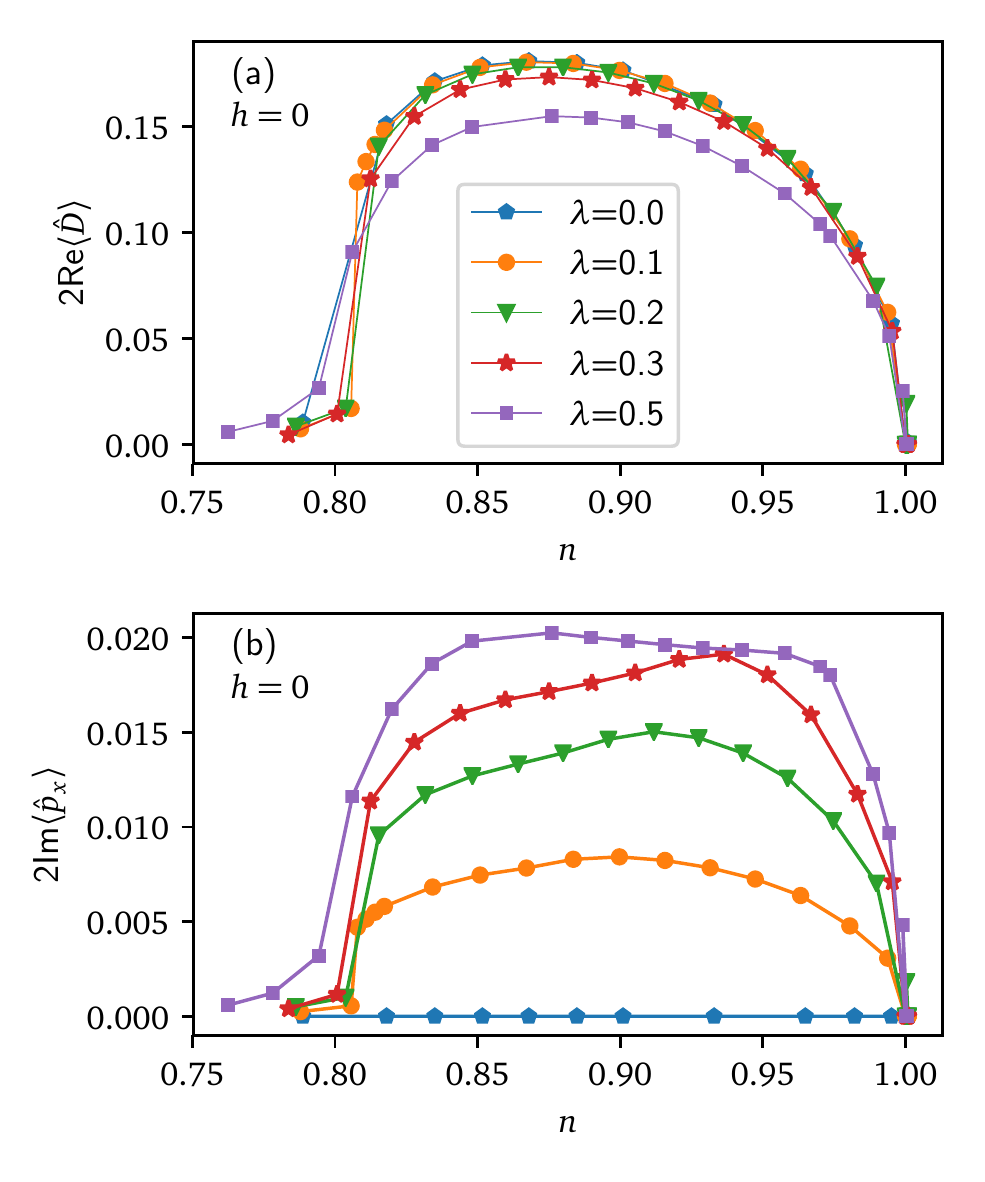}
\caption{(Color online). (a) $d$-wave and (b) $p$-wave order parameters as a function of density $n$ for various values of
spin-orbital coupling $\lambda$. 
The Zeeman magnetic field $h$  is zero and the interaction $U$ is set to $8.0$. 
Symbols in the lower panel have the same meaning as in the upper panel.}\label{fig:OP_vs_n_h0}
\end{figure}
\begin{figure}[h]
\includegraphics[width=0.94\columnwidth]{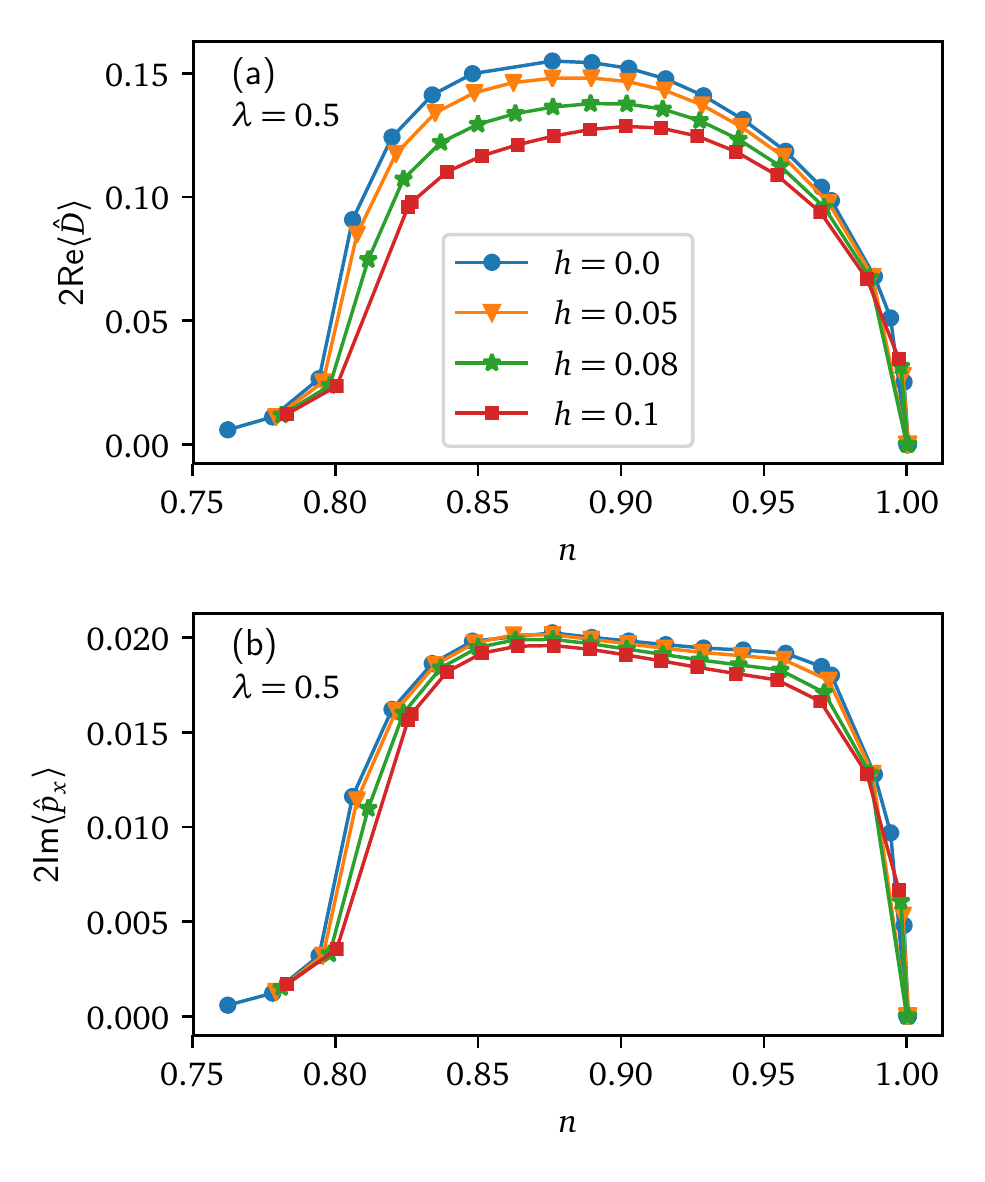}
\caption{(Color online). (a) $d$-wave and (b) $p$-wave order parameters as a function of density $n$ for various values of Zeeman magnetic field $h$. The spin-orbital coupling and interaction are set to $\lambda=0.5$ and $U=8.0$. 
Symbols in the lower panel have the same meaning as in the upper panel.}\label{fig:OP_vs_n_lambda05}
\end{figure}
\begin{figure*}[t]
\includegraphics[scale=0.8]{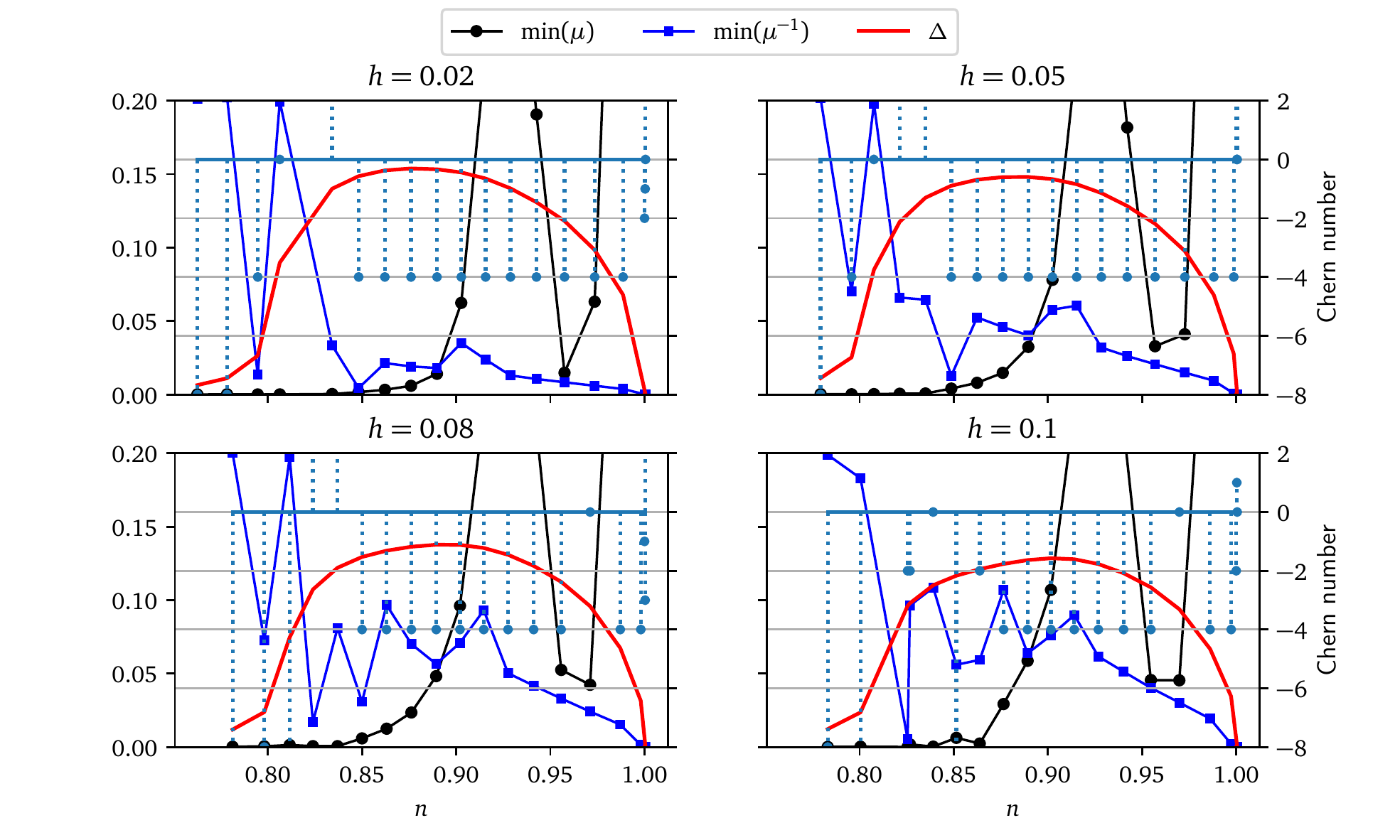}
\caption{(Color online).
Analysis of the topological character of the solutions illustrated on Fig.~\ref{fig:OP_vs_n_lambda05}.
For each of four values of the Zeeman field $h$, we show (i) the lower bound (or half gap) of the eigenvalue of the effective topological Hamiltonian $h_t(\kv)$ across the Brillouin zone; (ii) the lower bound of the inverse eigenvalue; (iii) the $d$-wave order parameter, for reference and (iv), with the scale appearing on the right, the computed Chern number. We see signs of a topological to nontopological transition within the superconducting phase, at around $n=0.84$ for $h=0.02$, 0.05 and 0.08 and around $n=0.86$ for $h=0.1$.
}\label{fig:gaps}
\end{figure*}

In practice, the numerical computation of the Chern number $C_1$ is done using the method proposed in Ref.~\cite{fukui_chern_2005}, which yields quantized (integer) values up to double precision accuracy.
However, blindingly following this procedure is dangerous; let us explain.
Expression \eqref{eq:Chern} yields a topological invariant only if the integral is carried over the full Brillouin zone with a continuous integrand. This is no longer true if the region where $\mu_{1,2}(\kv)<0$ does not cover the whole Brillouin zone, i.e., if $\mu_{1,2}(\kv)$ changes sign somewhere.
This can happen in two ways:
\begin{enumerate}
\item $\mu_{1,2}(\kv)$ goes through zero. There is a `Fermi surface', and the energy gap has closed.
\item There is a pole in the self-energy, i.e., a zero in the Green function. Then $\mu_1(\kv)$ goes to $-\infty$ and there is a `surface of zeros' in the Brillouin zone. The Wang-Zhang argument~\cite{wa.zh.12a, wa.zh.12b} explicitly excludes that case.
\end{enumerate}
If the system were normal (no superconductivity), these circumstances would lead to a nonquantized value of Expression \eqref{eq:Chern}.
But nothing prevents us from using the Nambu formalism even when the system is normal. 
If this is done, then the doubling of bands still leads to a quantized Chern number, because the region of the Brillouin zone where $\mu_{1,2}(\kv)<0$ is then perfectly complemented by the region where $\mu_{3,4}(\kv)<0$ (time-reversal symmetry breaking does not come from the momentum-dependent part of the Hamiltonian), and the sum of these two contributions, coming from opposite eigenvalues of the Nambu doubled Green function, yields an integer. However, this is clearly wrong: a blind computation of \eqref{eq:Chern} in the Nambu formalism will always produce a quantized result and is not enough to guarantee a topological state. 
A stringent criterion for the topological character of the solution is that  $\mu_{1,2}(\kv)$ does not change sign anywhere in the Brillouin zone.
Accordingly, we have numerically computed, from the periodized Green functions \eqref{eq:Gper} obtained from our solutions, the minimum $\min(\mu)$ of the absolute value  of the eigenvalues of $h_t(\kv) = - \mathcal{G}^{-1}(\kv,0)$ across the Brillouin zone, as well as the minimum inverse absolute value $\min(\mu^{-1})$. The system is topological only if (i) these minima are nonzero and (ii) the computed Chern number is nonzero.

\section{Results and discussion}\label{sec:res}
\subsection{the $d$-wave and $p$-wave parity-mixing phase}
In order to see how the Rashba spin-orbit coupling affects the superconducting state, we plot, on Fig.~\ref{fig:OP_vs_n_h0}, the $d$-wave and $p$-wave order parameters defined in~\eqref{eq:OP} as a function of the density, for various values of the Rashba coupling $\lambda$. 
When $\lambda=0.0$ and $h=0.0$, Hamiltonian~\eqref{RH} reduces to the pure Hubbard model and there is a $d$-wave superconducting solution in the hole-doped region, in agreement with previous DMFT studies~\cite{Capone2006, Kancharla2008}.
The $p$-wave order parameter then vanishes, as shown in Fig.~\ref{fig:OP_vs_n_h0}b. 
When the Rashba coupling is turned on, the $d$-wave order parameter is slightly suppressed and the $p$-wave order parameter increases from zero as a function of $\lambda$.
There is therefore a parity-mixing superconducting state in the region $0.8<n<1.0$.
This is the main result of this paper.
The superconducting solution has a phase freedom, which we fixed by imposing real values on the singlet pairing amplitudes  $\Delta_1$ and $\Delta_2$ defined in Eq.~\eqref{eq:bath_params}.
This leaves open the question of the phase of the triplet amplitudes. 
The choice made in Eq.~\eqref{eq:bath_params} (with $p_{x,y 1,2}$ real) produces the solution illustrated here.
On the other hand, choosing pure imaginary values for the same parameters converges to a non superconducting solution.
The solutions obtained with the convention \eqref{eq:bath_params} yield real values of the order parameter $\langle\hat D\rangle$ and imaginary values of $\langle\hat p_{x,y}\rangle$.
The effect of a Zeeman magnetic field on superconductivity is shown in Fig.~\ref{fig:OP_vs_n_lambda05}, where the strength of the Rashba coupling is fixed at $\lambda=0.5$, in order to allow the appearance of $p$-wave superconductivity. 
The Zeeman field suppresses both the $d$-wave and $p$-wave order parameters, but the effect is much smaller on the latter, because of the singlet-breaking effect of the Zeeman field.
\subsection{Topological properties}
An important role of the Zeeman magnetic field is to break time-reversal symmetry in the system, which leads to nontrivial
topological properties~\cite{sa.ta.09,yo.ya.16}.
Fig.~\ref{fig:gaps} shows our analysis of the topological properties of the solutions illustrated on Fig.~\ref{fig:OP_vs_n_lambda05}.
As explained in Section \ref{sec:topo}, a solution has a nontrivial topology only if (i) the computed Chern invariant is nonzero and (ii) the effective topological Hamiltonian is gapped and has quasiparticles, i.e., if its eigenvalues $\mu_\alpha(\kv)$ do not go to zero or to infinity within the Brillouin zone. The black curves on Fig.~\ref{fig:gaps} show the minimum value of $|\mu_\alpha(\kv)|$ in the Brillouin zone, as a function of electron density. The blue curves, on the other hand, show the minimum value of $|\mu^{-1}_\alpha(\kv)|$. Both these values need to be nonzero in order for the system to have nontrivial topology. The quantized Chern invariant computed from the methods of Refs~\cite{wa.zh.12a, wa.zh.12b, fukui_chern_2005} is shown by vertical impulses and is equal to $-4$ in the region where the conditions for nontrivial topology are satisfied. The dSC order parameter is shown in red for reference.
The most striking feature of these results is the apparent topological transition in the overdoped region of the phase diagram.
At around $n=0.84$, while the system is still superconducting, the gap in the eigenvalues of the effective topological Hamiltonian $-\mathcal{G}(\kv,0)$ seems to disappear. 
This coincides with erratic values of the numerically computed Chern number. We cannot rule out that this is a numerical effect, i.e., that the gap in the topological Hamiltonian is simply very small without being actually zero. Thus the gap-protected topology would simply become more and more fragile as doping is increased, with a rather sharp turn just after optimal doping.
It is also worth noting that the computed Chern number becomes erratic as half-filling is reached, which is explained by the appearance of zeros in  $\mathcal{G}(\kv,0)$, as expected in a Mott insulator.
\section{Conclusions}\label{sec3}
We have studied a Hubbard model with Rashba spin-orbit coupling using cluster dynamical mean field theory and have observed the appearance of mixed-parity superconductivity away from half-filling. 
Increasing the Rashba coupling moderately suppresses the $d$-wave order parameter but greatly enhances the associated $p$-wave order parameter.
Adding a Zeeman field breaks time-reversal symmetry and gives the superconducting state a topological character.
The Chern number can be computed in the bulk and is equal to $-4$ in a wide range of doping.
The topological character is dramatically weakened at about 15\% doping, just a little after optimal doping.
Whether this is a continuous phenomenon or a well-defined, doping-driven topological transition is not absolutely clear from the numerical results.

One possibility is that this behavior in the overdoped region is the signature of a first-order transition between two competing CDMFT solutions. This could be checked by sweeping the procedure finely as a function of chemical potential away and towards half-filling. Resource and time limitations did not allow this to be done.
Another possible improvement on the results presented here would be to greatly increase the number of CDMFT parameters, which we have kept at a minimum in order to save time and facilitate convergence.

Another way to detect the topological character of the system is to study a ribbon geometry and to detect edge states.
This has been done at the mean-field level in Ref.~\cite{yo.ya.16}.
In CDMFT, this would require to align up a fair number of independent, one-dimensional systems in, say, the $y$ direction, and to solve the combined system in a real-space CDMFT framework~\cite{charlebois_impurity-induced_2015}.
One would then need to probe the interacting spectral function in order to detect edge modes; see, for instance, Fig.~8 of~\cite{hassan_quarter-filled_2013}.
This procedure is rather resource-intensive, because of the multiple copies of the impurity system needed for each set of parameters, and is deferred to further work.

\begin{acknowledgments}
Discussions with A. Foley are gratefully acknowledged.
Computing resources were provided by Compute Canada and Calcul Qu\'ebec. 
X.L. is supported by the scholarship from China Scholarship Council (CSC) and the Fundamental Research Funds for Central Universities (No. 20720180015). His stay at Universit\'e de Sherbrooke was supported in part by the \textit{Institut quantique}.
\end{acknowledgments}
\bibliographystyle{apsrev}

\end{document}